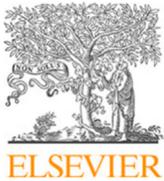
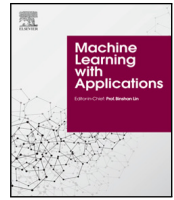

# DAIRE: A lightweight AI model for real-time detection of Controller Area Network attacks in the Internet of Vehicles

Shahid Alam [a],[*], Amina Jameel [b], Zahida Parveen [a], Ehab Alnfrawy [a], Adeela Ashraf [c], Raza Uddin [d], Jamal Aqib [e]

[a] *Department of Information Security, College of Computer Science and Engineering, University of Ha'il, Ha'il, Saudi Arabia*
[b] *Department of Computer Engineering, Center of Excellence in Artificial Intelligence, Bahria University, Islamabad, Pakistan*
[c] *Department of Artificial Intelligence, College of Computer Science and Engineering, University of Ha'il, Ha'il, Saudi Arabia*
[d] *Department of Computer Science, College of Computer Science and Engineering, University of Ha'il, Ha'il, Saudi Arabia*
[e] *Department of Software Engineering, College of Computer Science and Engineering, University of Ha'il, Ha'il, Saudi Arabia*

A B S T R A C T

The Internet of Vehicles (IoV) is advancing modern transportation by improving safety, efficiency, and intelligence. However, the reliance on the Controller Area Network (CAN) introduces critical security risks, as CAN-based communication is highly vulnerable to cyberattacks. Addressing this challenge, we propose DAIRE (Detecting Attacks in IoV in REal-time), a lightweight machine learning framework designed for real-time detection and classification of CAN attacks. DAIRE is built on a lightweight artificial neural network (ANN) where each layer contains $N_i = i \times c$ neurons, with $N_i$ representing the number of neurons in the $i$th layer and $c$ corresponding to the total number of attack classes. Other hyperparameters are determined empirically to ensure real-time operation. To support the detection and classification of various IoV attacks, such as *Denial-of-Service*, *Fuzzy*, and *Spoofing*, DAIRE employs the sparse categorical cross-entropy loss function and root mean square propagation for loss minimization. In contrast to more resource-intensive architectures, DAIRE leverages a lightweight ANN to reduce computational demands while still delivering strong performance. Experimental results on the CICIoV2024 and Car-Hacking datasets demonstrate DAIRE's effectiveness, achieving an average detection rate of 99.88%, a false positive rate of 0.02%, and an overall accuracy of 99.96%. Furthermore, DAIRE significantly outperforms state-of-the-art approaches in inference speed, with a classification time of just 0.03 ms per sample. These results highlight DAIRE's effectiveness in detecting IoV cyberattacks and its practical suitability for real-time deployment in vehicular systems, underscoring its vital role in strengthening automotive cybersecurity.

## 1. Introduction

The Internet of Things (IoT) has revolutionized connectivity by integrating smart devices into a unified network. As a developing technology, IoT aims to integrate computing intelligence into in-vehicle systems to efficiently manage driving conditions (Tao et al., 2019). The automotive industry is experiencing a paradigm shift with the emergence of the Internet of Vehicles (IoV), where vehicles are increasingly connected to each other, to infrastructure, and to cloud-based services. IoV facilitates communication not only between vehicles but also between vehicles and external entities such as infrastructure, pedestrians, and other smart devices (Mehedi et al., 2021). This connectivity enhances traffic safety, route optimization, and autonomous driving capabilities. However, the increasing reliance on interconnected systems also exposes vehicles to significant security and privacy risks, making IoV a prime target for cyberattacks (Lydia et al., 2021).

IoV systems consist of both intra-vehicle and inter-vehicle communication networks. With the growing number of smart and connected vehicles, these networks face escalating security risks (Ullah et al., 2024). Vehicles are equipped with multiple electronic control units (ECUs) that manage functions like braking, steering, and acceleration. These ECUs typically communicate over the Controller Area Network (CAN), a widely used communication protocol in vehicles that facilitates message exchange among ECUs (Song et al., 2020). Despite its efficiency and real-time performance, the CAN protocol was not originally designed with security in mind. It lacks authentication and encryption, making it vulnerable to injection, spoofing, and denial-of-service (DoS) attacks (Yang et al., 2023). Attackers can manipulate vital vehicle functions by forging CAN messages, potentially jeopardizing vehicle safety and passenger security.

In addition to vulnerabilities in intra-vehicle networks, IoV's broader communication environment also introduces multiple security challenges (Al-Jarrah et al., 2019). Each connected node, whether a sensor, vehicle, or infrastructure component, serves as a potential entry point for cyber threats. Traditional security measures have proven inadequate for detecting sophisticated and evolving cyberattacks in such environments. Therefore, advanced techniques are required to strengthen the overall security of IoV networks (Seo et al., 2018).

Recent research has increasingly focused on machine learning (ML), one of the main branches of AI, approaches for enhancing intrusion detection systems (IDS) in IoV (Li et al., 2023; Mishra et al., 2018). ML-based IDS has shown promise in identifying known attacks, but it struggles with unknown threats. To address this limitation, researchers have turned to ensemble learning methods, which combine multiple classifiers to improve detection accuracy and robustness (Yazdinejad et al., 2023). These models use feature engineering and meta-learning (stacking and voting techniques) to enhance classification performance, making them highly effective for securing IoV networks. Given the constraints of modifying legacy in-vehicle systems, lightweight IDS solutions that detect anomalies without adding overhead are essential (Song et al., 2020).

Deep Learning (DL), a subset of ML, models have demonstrated superior capabilities in learning data representations and identifying anomalous patterns (Tangade et al., 2019). These methods excel at identifying complex patterns in large volumes of data and can be trained to detect subtle deviations from normal behavior (LeCun et al., 2015). This helps identify cyber-attacks in real-time. Recent works have explored neural networks, including convolutional and recurrent architectures, to model temporal dependencies and classify anomalous CAN messages. However, challenges remain in achieving low-latency, high-accuracy detection that is practical for real-world deployment in resource-constrained vehicular environments (Yang et al., 2022a).

This paper presents a lightweight ML-based detection model aimed at identifying CAN bus attacks in IoV environments in real-time. The proposed approach uses the power of neural networks to provide robust, scalable, and adaptive protection against cyber threats targeting vehicular communication systems.

The key contributions of this work are as follows:

- **DAIRE Model**: A lightweight ML model for real-time classification of CAN bus attacks in IoV environments. The model employs a lightweight Artificial Neural Network (ANN) architecture.
- **Lightweight Design**: Introduces a lightweight ANN structure where the number of neurons in each layer is computed as $N_i = i \times c$ (where $N_i$ is neurons in the $i$th layer and $c$ is the number of classes). This design ensures computational efficiency while maintaining high accuracy.
- **Efficient Loss Function and Optimizer**: The use of Sparse Categorical Cross-Entropy (SCCE) for multinomial classification (memory-efficient compared to CCE) and RMSprop for adaptive learning rate optimization, enhancing convergence speed.
- **Performance**: The model achieves an average detection rate (DR) of 99.88%, false positive rate (FPR) of 0.02%, and accuracy of 99.96% on benchmark datasets CICIoV2024 and Car-Hacking. It outperforms existing methods with a classification time of 0.03 ms per sample.
- **Comprehensive Evaluation**: The model is validated on diverse attack types using real-world datasets, demonstrating robustness and generalizability.

The remainder of this paper is organized as follows. Section 2 provides background on ANN architecture, and it also explains CAN bus vulnerabilities. Section 3 reviews related works in learning based intrusion detection for IoV. Section 4 explains the design and implementation of the proposed DAIRE model. Section 5 presents the experimental evaluation and results. Section 6 discusses comparative analysis and future directions. Finally, Section 7 concludes the paper.

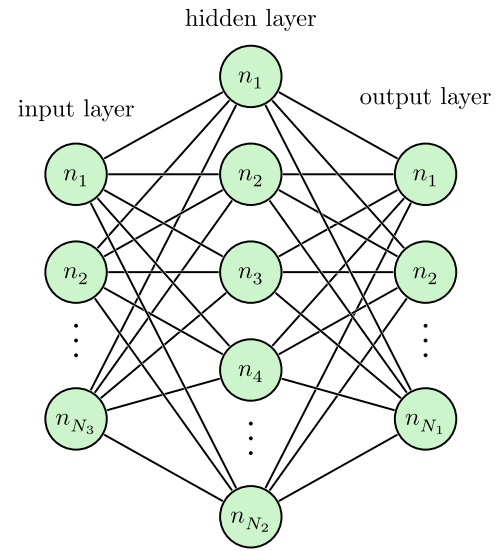

**Fig. 1.** A basic ANN architecture.

## 2. Background

As vehicles evolve into intelligent and connected systems, the role of cybersecurity in vehicular networks has become increasingly critical (Al-Jarrah et al., 2019). IoV extends the concept of IoT to transportation by enabling continuous interaction between vehicles, roadside infrastructure, and cloud services. This advancement improves road safety, traffic efficiency, and driving experience, but also exposes the system to new cybersecurity vulnerabilities.

To understand DAIRE's (Detecting Attacks in IoV in REal-time) design, we briefly examine the core structure of ANNs and the CAN protocol. These fundamental components will help to understand our IDS that addresses IoV security challenges through lightweight computations and real-time analysis.

### 2.1. Artificial Neural Network (ANN)

ANNs are a class of ML models inspired by the structure and functioning of the human brain. They consist of interconnected layers of nodes or *neurons* that process input data and learn patterns through weighted connections. Typically, an ANN includes an input layer, one or more hidden layers, and an output layer (Shanmuganathan, 2016). Each neuron processes its inputs using an activation function and passes the output to the subsequent layer. The basic architecture of an ANN is shown in Fig. 1. There are three layers in this ANN. The number of neurons in the output layer, hidden layer, and input layer are $N_1$, $N_2$, and $N_3$, respectively. A basic ANN typically has one hidden layer (as shown in Fig. 1), i.e., it is a shallow neural network, while deeper networks have many.

ANNs are particularly useful for classification tasks, as they can approximate complex functions and recognize non-linear relationships in data. Their simplicity and computational efficiency make them especially suitable for real-time and embedded applications, such as intrusion detection in vehicular networks. Compared to Convolutional Neural Networks (CNNs), which are specialized for image and spatial data, ANNs are lightweight and require fewer computational resources.

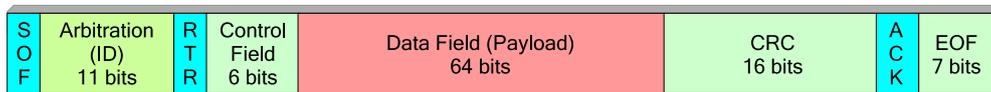

**Fig. 2.** A basic CAN packet structure. SOF and RTR are 1 bit each. ACK is 2 bits. The ID bit is 11 bits in the base format and 29 bits in the extended format.

This makes them ideal for processing structured data, such as CAN messages, where the input is numeric and does not exhibit spatial hierarchies.

The core of an ML model is the ANN employed in the model. ANNs are designed to work like human brains. They have neurons interconnected with each other through layers that process information and learns by extracting patterns from the data. The input is processed by the network, and an output is generated. A loss function calculates the difference between the predicted and the true output. The loss is propagated back to the network with adjusted parameters, e.g., updated weights, to minimize the loss.

In the context of IoV, the use of ANNs enables the detection and classification of cyberattacks based on the structure and timing of CAN messages. By adjusting weights using backpropagation and optimization algorithms like RMSprop, ANNs can be trained to distinguish between normal and malicious communication patterns with high accuracy and low latency. The lightweight and efficient nature of ANNs makes them particularly suitable for processing structured data like CAN messages. This aligns with the need for real-time intrusion detection in resource-constrained vehicular environments.

In addition to basic ANNs, advanced architectures such as CNNs, Recurrent Neural Networks (RNNs), and Long Short-Term Memory (LSTM) networks have also been developed to address complex data types and temporal patterns. CNNs are particularly effective for spatial data like images and are widely used in computer vision, but they are computationally intensive due to their deep hierarchical structure. While they could theoretically process CAN traffic as a 1D sequence, their overhead is often unjustified for the relatively simpler patterns in CAN attack detection. RNNs and LSTMs are tailored for sequential data, capturing temporal dependencies in time-series applications such as speech recognition or CAN bus traffic analysis. These models capture long-range dependencies but suffer from high memory usage and latency due to their recursive nature. Though CAN bus traffic is time-series data, the shorter sequential dependencies in attack signatures may not require LSTM-level complexity.

Despite their powerful learning capabilities, these architectures often require substantial computational resources and memory, making them less suitable for resource-constrained environments such as embedded automotive systems. In contrast, the DAIRE model utilizes a lightweight ANN to ensure low-latency, lightweight deployment in real-time IoV scenarios. While deeper models like CNNs and LSTMs offer performance advantages in some domains, our empirical evaluation demonstrates that a lightweight ANN can achieve competitive accuracy in CAN attack detection without the overhead of complex architectures. This trade-off between complexity and efficiency makes our ANN-based DAIRE model a practical choice for IoV environments.

### 2.2. Controller Area Network (CAN)

CAN protocol is a widely adopted communication standard used in vehicles to enable message exchange between ECUs (Song et al., 2020). CAN supports real-time, high-speed, and reliable in-vehicle communication across components such as the engine control module, brake system, and steering mechanisms. Despite its widespread adoption, the CAN protocol lacks fundamental security features such as encryption, authentication, and access control. This makes it inherently vulnerable to several types of cyberattacks (Yang et al., 2023), including:

- *Denial-of-Service (DoS)* attacks involve an attacker overwhelming the CAN bus with high-priority messages, effectively preventing critical commands from being transmitted. This can disrupt or completely disable communication between legitimate ECUs, potentially leading to partial or total failure of vehicle systems, including vital safety features.
- *Spoofing* attacks occur when an attacker sends counterfeit or modified CAN messages, mimicking legitimate ECUs. The intent is to trick other ECUs into accepting these messages as authentic, which can result in unsafe or unintended vehicle behavior. There are several different kinds of spoofing attacks:
    - *Speed Spoofing*: Inserting fake CAN messages or modifying genuine ones to display an incorrect vehicle speed. This misleads the driver and creates a false perception of velocity.
    - *Steering Wheel Spoofing*: Tampering with, falsifying, or overriding steering-related CAN signals. This disrupts functions such as lane keeping, lane changing, and trajectory control.
    - *Gas Spoofing*: Generating forged CAN frames that misrepresent pedal position, e.g., showing full throttle when the driver is not pressing the pedal. This may result in sudden acceleration or a lack of expected acceleration.
    - *RPM Spoofing*: Injecting fabricated engine speed (RPM) data into the CAN network, causing the dashboard to display incorrect values. This may interfere with safety mechanisms like automatic transmission control.
    - *Gear Spoofing*: Sending counterfeit CAN messages about gear position, e.g., reporting Reverse while the car is actually in Drive. This can lead to autonomous functions operating in the wrong direction.
- *Fuzzy* attacks consist of sending random CAN IDs and data payloads into the network in an attempt to crash or destabilize it. This method is also used to uncover potential weaknesses in the embedded software.

These vulnerabilities arise mainly because the original CAN specification prioritized efficiency and fault tolerance over security. As a result, attackers can exploit the open nature of CAN communication to manipulate vehicular functions (Yang et al., 2023). Securing CAN communication in IoV scenarios is a critical priority. Recent research efforts have focused on identifying and mitigating vulnerabilities within the CAN bus system, particularly concerning intra-vehicle security. The main aim is to address the protocol's security weaknesses from multiple perspectives (Aliwa et al., 2021). Given these security limitations, understanding the structure of CAN packets is essential to identifying how attackers exploit specific fields

### 2.3. CAN packet structure

Visualizing the CAN message format helps in understanding how attacks manipulate standard messages to achieve malicious objectives. Fig. 2 illustrates the structure of a CAN packet, which begins with a Start of Frame (SOF) and ends with an End of Frame (EOF), occupying 1 bit and 7 bits, respectively. The arbitration field is an 11-bit identifier that determines which node gains control of the bus. It plays a critical role in determining the priority of messages on the bus. The arbitration field includes a bit for Remote Transmission Request (RTR). The Control Field (6 bits) specifies data length, followed by the Data

Field (Payload), which can carry up to 64 bits of payload. A 16-bit Cyclic Redundancy Check (CRC) ensures message integrity by detecting transmission errors in the CAN frame, while the ACK field (2 bits) confirms successful receipt.

The lack of authentication and encryption in the CAN protocol significantly exposes the network to various cyber threats (Yang et al., 2023). Attackers can exploit these weaknesses by injecting fabricated messages with forged *Arbitration IDs*, effectively spoofing legitimate ECUs on the network. DoS attacks may be carried out by overwhelming the bus with a stream of high-priority frames, especially those containing dominant 0 bits in the Arbitration Field, thereby obstructing normal communication. Attackers tamper with the *Data Field* to change critical payloads, such as changing speedometer readings or sensor values, posing serious risks to vehicle functionality and passenger safety.

## 3. Related works

The IoV relies heavily on the CAN bus, which inherently lacks security mechanisms, rendering it susceptible to cyber threats such as spoofing and DoS attacks. These vulnerabilities arise due to the CAN protocol's transmission of unencrypted messages (Aswal & Pathak, 2024; Lee et al., 2021). Recent research has increasingly explored ML approaches for real-time intrusion detection in CAN-based systems. This section reviews recent studies that have proposed ML models aimed at identifying cyber attacks within the IoV context.

DL models have demonstrated superior capabilities in learning complex patterns from CAN bus data. For instance, Song et al. (2020) proposed a deep convolutional neural network (Deep CNN) based IDS for CAN bus networks in vehicles. It uses a reduced Inception-ResNet architecture. The method utilizes raw bitwise CAN ID sequences converted into 2D frames, enabling the model to learn sequential patterns without requiring manual feature engineering. The model outperformed conventional ML approaches, achieving high accuracy and F1 score across multiple attack types, including DoS, spoofing, and fuzzy attacks. The method processed real CAN data in near real-time with minimal error, but it still relies on supervised learning, limiting the detection of unknown attacks. The work by Sreelekshmi and Aji (2025) introduced a deep learning architecture that models CAN traffic as CANGraph-feature images. Their method transforms message interactions into graph structures and subsequently into images to capture complex structural and contextual patterns, which are then classified using a CNN. This approach aims to detect subtle anomalies by using the spatial feature extraction power of CNNs.

Hanselmann et al. (2020) presented CANet, an unsupervised IDS for high-dimensional CAN bus data, which uses DL to detect abnormalities in the absence of labeled data, demonstrating excellent generalizability to various attack types. CANShield, a signal-level IDS for CAN developed in Shahriar et al. (2022) employs several auto-encoders and identifies stealthy intrusions more efficiently by evaluating signal behavior at the physical layer. A BERT language model is used in Alkhatib et al. (2022) to capture contextual links between CAN message sequences for high-accuracy detection. Ulmer (2023) proposed a DL-based IDS that protects CAN networks by detecting traffic irregularities using temporal and payload-based features.

Refat et al. (2021) extracted graph-based features from CAN data and used traditional ML algorithms to detect anomalies. Korium (2024) developed a secure and efficient DL framework targeted at latency-sensitive automotive systems. Yanmin et al. (2025) presented a hybrid LSTM-GRU architecture for detecting both known and zero-day intrusions, leveraging temporal dependencies in CAN data. In Xing et al. (2023), a spatiotemporal feature-based IDS was created to capture time series patterns and contextual correlations.

Federated learning (FL) has also emerged as a promising solution for preserving data privacy while enabling collaborative learning. Yang et al. (2022b) introduced an FL-based in-vehicle network intrusion detection system (IVN-IDS) for the IoV. They used a ConvLSTM model, which was trained locally across connected vehicles. Their framework incorporates a PPO-based client selection algorithm to enhance convergence, minimize overhead, and maintain model accuracy. It was trained on a real CAN bus dataset. The system achieved an average detection rate of 90.6% and an accuracy of 94%. However, the system was not able to accurately identify drop attacks and adapt to unseen threats.

Huang et al. (2024) presented an FL-based IDS that employs MobileNet-Tiny to provide precise, real-time detection on edge devices. Xie et al. (2024) proposed IoV-BCFL, an intrusion detection framework that integrates FL with blockchain technology. The FL component enables distributed model training across vehicle nodes and aggregation at Road Side Units (RSUs), which reduces communication overhead and preserves data privacy. In addition, the blockchain-based logging mechanism utilizes RSA encryption and InterPlanetary File System (IPFS) to securely store and manage intrusion records. The system employs smart contracts to log detected attacks. This facilitates intruder tracking, vulnerability analysis, and forensic evidence collection. The results on public datasets showed that IoV-BCFL achieves high detection rates while maintaining efficient performance in terms of communication latency, throughput, and smart contract reliability. Althunayyan et al. (2024) proposed a hierarchical federated learning framework that allows for decentralized and privacy-preserving IDS training. Wang et al. (2023) investigated a multi-view graph learning technique in their StatGraph model, which improves the adaptability of IDS in in-vehicle networks. Wang et al. (2025) proposed ATHENA, a hybrid intrusion detection system that combines physical-layer signals with semantic features, to boost resilience against hostile attacks.

DL based methods improve detection accuracy by learning from heterogeneous data, but they often overlook behavioral data analysis and raise privacy risks due to centralized data sharing. Chen et al. (2024) presents a behavior-aware intrusion detection model called FDL-IDM. It combines FL with differential privacy. It processes driving behavior data both spatially and temporally, uses attention mechanisms to enhance sequence modeling, and applies noise perturbation to protect privacy during training. The limitations include reliance on synthetic datasets, noise sensitivity, and scalability concerns in resource-constrained settings.

In order to balance interpretability and performance, Cheng et al. (2023) presented LSF-IDM, a lightweight model that integrates attribution and semantic fusion. Sousa et al. (2023) developed a lightweight IDS designed for 5G-enabled IoV environments, with a focus on lowering latency while retaining detection performance. Palma et al. (2025) investigated class imbalance in IDS data by testing and optimizing multiple ML classifiers for real-world IoV deployments. Furthermore, Devnath (2023) introduced GCNIDS, a novel graph convolutional network-based method for identifying structural relationships in CAN traffic and enhancing detection accuracy for complicated intrusion patterns.

To address the limitations of single-model approaches, researchers have turned to ensemble and hybrid methods. Alladi et al. (2021) suggested a DL-based IDS that uses LSTM and CNN models on edge devices to overcome the delay in IoV intrusion detection. Fu et al. (2025) introduced IoV-BERT-IDS, a hybrid intrusion detection framework that uses BERT to process both in-vehicle and extra-vehicle network traffic. The authors proposed a novel semantic extractor to convert raw traffic into contextual byte sentences and employ masked and next-sentence prediction tasks to enhance feature learning. The model was trained on data sets such as CICIDS, Car Hacking, and IVN-IDS and achieved an average accuracy of 99.98% and an $F_1$ score of 0.9998. Despite its effectiveness and strong generalization, especially across vehicle types, the framework faced challenges in lightweight deployment and detecting rare attack classes. Ullah et al. (2025) developed a hybrid ensemble model for IoV intrusion detection. The model combined XGBoost, Random Forest, and other classifiers with SMOTE-based imbalance handling and Information Gain-driven feature selection. The approach achieved 99.74% accuracy on CIC-IDS2017 and 100% on UNSW-NB15. However, the model's reliance on public

**Table 1**
Comparison of Literature Review on detection of cyber attacks against IoV.

| Article | Dataset | DR | Accuracy | $F_1$-score | Key features | Research gaps |
| --- | --- | --- | --- | --- | --- | --- |
| Song et al. (2020) | Car-Hacking | 99.84% | 99.92% | 99.91% | CAN data to image; Deep CNN; Adapted Inception-ResNet. | Real-time deployment may face challenges in resource-constrained embedded systems. |
| Alladi et al. (2021) | VeReMi Extension | 98.26% | 99.65% | 98.26% | MEC-based; Time-sequence and image-based classification. | All misbehavior types are not included. |
| Yang et al. (2021) | CAN-Intrusion; CICIDS-2017 | 99.99% 99.81% | 99.99% 99.87% | 99.99% 99.87% | Information gain; Fast correlation; PCA; Multitiered; Stacking; $k$-means clustering | Complex data distributions; Extensive preprocessing and feature tuning. |
| Yang et al. (2022b) | HCRL | 90.63% | 94.05% | 91.15% | ConvLSTM; Federated Learning (FL); Proximal Policy Optimization (PPO) | Computational complexity of PPO and FL. |
| Huang et al. (2024) | Car-Hacking; CICIDS2017 | 99.21% 98.76% | 99.17% 98.86% | 99.1% 98.9% | CAN data to image; FL; MobileNet-Tiny. | No formal privacy-proof against gradient leakage; Relies on a semi-honest model. |
| Qin et al. (2024) | Proprietary | 99.61% | 96.24% | 96.32% | Cloud-based training; Multidimensional features; XGBoost. | Prone to adversarial attacks; May not scale well. |
| Chen et al. (2024) | CAN-IDS; UNSWNB-15; CICIDS-2017 | 99.79% 97.33% 97.98% | 98.56% 96.49% 96.94% | 98.51% 97.51% 97.89% | FL; Differential privacy; Behavior analysis in temporal and spatial Dimensions. | Noise in differential privacy can degrade model accuracy. |
| Xie et al. (2024) | UNSW NB15; Car-Hacking | 99.97% 99.75% | 99.97% 99.83% | 99.97% 99.82% | Privacy-Preserving FL; Blockchain Integration. | Scalability concerns with blockchain overhead. |
| Ullah et al. (2025) | CICID2017; UNSW-NB15 | 99.74% 100% | 99.74% 100% | 99.74% 100% | Ensemble-based model; Information Gain for feature selection. | Zero-day attacks or highly adaptive adversaries have not been evaluated. |
| Fu et al. (2025) | CICIDS-2017; Car-Hacking; IVN-IDS | 100% 99.98% 99.96% | 99% 99.97% 99.96% | 100% 99.85% 99.96% | Semantic extraction (SE) and BERT; SE for converting traffic data into byte sentences. | High computational load of BERT makes real-time deployment challenging. |

datasets and computational demands may limit scalability for real-time IoV deployments.

While most of the above studies focus on intra-vehicle CAN security, other research has addressed higher-layer threats in Intelligent Transportation Systems. For example, Karthikeyan and Usha (2022) proposed an entropy-driven, reinforcement-learning-based approach for real-time DDoS flooding attack detection at RSUs in intelligent transportation systems. Their work demonstrates the feasibility of applying adaptive AI methods to mitigate large-scale network-level attacks. However, their model targets vehicle-to-infrastructure traffic rather than the in-vehicle CAN bus, and it employs comparatively heavier learning strategies that may not suit embedded ECUs. At the same time, efforts to improve efficiency have driven the development of new hybrid and lightweight approaches. For instance, recent work on DDoS detection (Karthikeyan & Usha, 2022) has highlighted the critical need for real-time responsiveness. Furthermore, models combining Autoencoders with Temporal Convolutional Networks aim to balance temporal feature learning with a compact architecture, while other studies explicitly propose lightweight hybrid frameworks for resource-constrained vehicles, underscoring the field's focus on this critical trade-off (Aljabri et al., 2025a). Luo et al. (2025) introduced a lightweight method using spectral residuals and a depthwise separable CNN that achieves high performance with low resource usage. Its effectiveness against dynamic attack patterns and its suitability for real-time deployment in diverse vehicular environments remain to be fully validated. Addressing a similar need for adaptable and efficient detection, Aljabri et al. (2025b) recently proposed a lightweight, data-driven IDS capable of multi-class intrusion detection across both in-vehicle and external network datasets. While their model demonstrates high accuracy and adaptability, its architecture is based on dense layers and the Adam optimizer.

A summary of several of these related studies is presented in Table 1, emphasizing their main features and identifying existing research gaps. The aforementioned studies demonstrate the growing effectiveness of ML and DL approaches for detecting CAN bus intrusions in IoV systems. Deep models like Deep CNNs, LSTM-GRU hybrids, and transformer-based architectures have shown high detection rates; however, many face limitations such as high computational complexity, reliance on extensive feature engineering, or dependence on centralized data that raises privacy concerns. Moreover, the real-time requirements and resource constraints of vehicular systems necessitate solutions that strike a balance between detection performance, interpretability, and computational efficiency.

To address these limitations, this paper introduces DAIRE, a lightweight and real-time machine learning-based IDS designed specifically for CAN bus security in IoV. Unlike heavy DL architectures, DAIRE employs a lightweight ANN structure that is both interpretable and optimized for low-latency environments.

## 4. DAIRE (detecting attacks in Iov in REal-time)

ML (LeCun et al., 2015) enables computational models with multiple processing layers to learn data representations at various levels of abstraction. An ANN uncovers complex patterns in large datasets using the backpropagation algorithm, which adjusts internal parameters to refine representations at each layer based on the previous layer.

ANN models use multi-layered neural networks for learning, and require a large amount of data and computational power to train. They are being applied in computer vision, self-driving cars, driver assistance systems, intrusion detection, biometric authentication, and real-time language translation etc. In real-time classification, ANN models analyze the incoming data instantly and categorize the data into predefined classes. This is crucial in applications that require immediate decision-making, such as autonomous driving and intrusion detection, etc. One such area is to detect attacks in real-time against CAN in IoV. There are different types of such attacks against IoV, such as DoS and Spoofing attacks. *DoS* attack against CAN can be launched by manipulating the arbitration mechanism. For example, flooding the CAN bus with misleading high-priority messages can prevent essential commands from passing through (Derhab et al., 2021). This can lead to dangerous consequences, such as disruption in the cruise control system during driving, and can cause a serious accident. In *Spoofing* attacks, the attacker takes over the CAN bus and acts as a legitimate node. For example, this can lead to controlling the Revolutions Per Minute (RPM) of the vehicle during driving and can cause a significant risk to the occupants (Taslimasa et al., 2023).

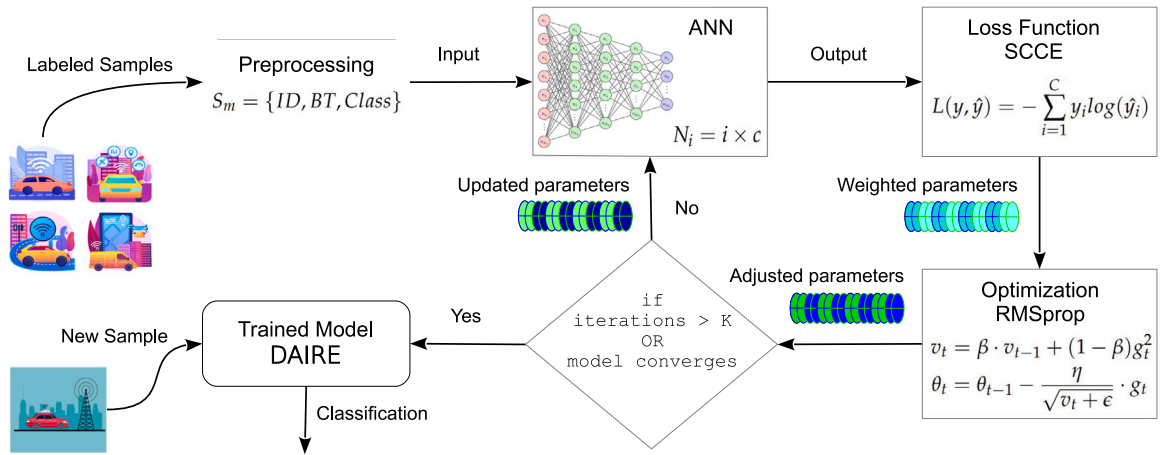

**Fig. 3.** A high-level overview of DAIRE. K represents the number of iterations used during optimization, commonly referred to as the number of *epochs*. It is typically assigned a large value between 500–1000. The ideal value of K is determined experimentally, based on the point at which the model converges, i.e., when its predictions no longer improve or the error rate stabilizes. For the dataset used in this paper, K = 100.

There is a need to develop real-time detection/classification of such attacks. In this paper, we propose, design, and develop a lightweight ML model named DAIRE (Detecting Attacks in Iov in REal time) that classifies such attacks in real-time. Fig. 3 presents a high-level overview of DAIRE. In the next few sections, we present the design and development of DAIRE.

### 4.1. Preprocessing

This section describes the data preprocessing, class imbalance problem, and feature extraction and normalization.

#### 4.1.1. Data preprocessing and class imbalance handling

The CICIoV2024 and Car-Hacking datasets have significant natural class imbalance, as detailed in Table 4. This reflects real-world IoV environments where normal CAN messages substantially outnumber attack instances. Rather than applying synthetic oversampling techniques like SMOTE (Synthetic Minority Over-sampling Technique) or GAN-based (Generative Adversarial Network-based) generation, we intentionally preserved the original data distribution because we aimed to evaluate DAIRE under realistic conditions where attack frequencies naturally vary. Artificially balancing the datasets might create an unrealistic performance expectation. In actual vehicular networks, the frequency of attacks is inherently imbalanced; normal messages vastly outnumber malicious ones. Training on the raw, imbalanced distribution ensures that DAIRE is optimized for the high-precision detection of rare attacks within a predominant stream of normal traffic, which is critical for real-world deployment.

Despite the imbalance, DAIRE achieved high performance across all classes, demonstrating its inherent capability to handle imbalanced data without artificial balancing. During experimental evaluation, our results demonstrate that DAIRE achieves high detection rates across minority attack categories without artificial balancing, confirming its inherent capability to learn effectively from imbalanced data.

#### 4.1.2. Feature extraction and normalization

The input to our ML model consists of data, as shown in Table 2 and Fig. 2, that the model uses to learn patterns, make predictions, and classify. This input consists of a set of input samples with features $S_m = \{ID, BT, Class\}$ in the form of CAN messages, where $m$ is the $m$th CAN message, $ID$ is the *Arbitration* field consisting of the identifier (ID), $BT = \{b_1, b_2, b_3, \ldots, b_n\}$ is the bytes transmitted comprising the *Data Field*, and *Class* is the specific class of the message. Raw hexadecimal CAN messages were converted into decimal feature vectors containing the Arbitration ID and Data Bytes, as illustrated in Table 3. All numerical features (ID and Data Bytes) were normalized to the [0, 1] range to ensure stable convergence during training. Attack classes were encoded as integer labels for efficient processing with the sparse categorical cross-entropy loss function. We have assigned different numbers to *Class* depending on the type of message. For example 0 if the message is normal, 1 if a DoS attack, 2 if a Fuzzy attack, 3 represents RPM Spoofing, and 4 is for SPEED Spoofing attack. An example of five samples of CAN messages belonging to five different classes is shown in Table 2.

We have preprocessed the labeled dataset. This preprocessing transforms raw CAN messages into structured feature vectors (ID, BT, Class) through normalization and categorical encoding. Two samples of CAN raw data with their corresponding feature vectors are shown in Table 3. These vectorized samples are now in a numerical format suitable for an ML model and serve as input features for our ANN architecture. Each feature vector preserves the temporal and payload characteristics necessary for attack classification.

### 4.2. Lightweight Artificial Neural Network (ANN)

In order to perform effective classification and enable real-time detection of IoV attacks, we have proposed a custom ANN architecture specifically designed to meet the constraints and demands of vehicular environments. Our model consists of an input layer, a few hidden layers, and an output layer. The number of hidden layers and other hyperparameters is computed empirically during the training of the model. We developed and computed the number of neurons in each layer using the following formula:

$$N_i = i \times c \qquad (1)$$

**Table 2**
Five samples of CAN messages belonging to five different classes.

```
125,000,000,247,128,000,063,255,255,0 => ID=125, Class=0 - Normal
291,001,005,009,007,000,011,015,004,1 => ID=291, Class=1 - DoS
643,218,183,114,074,036,000,112,006,2 => ID=643, Class=2 - Fuzzy
513,160,015,003,037,168,053,148,034,3 => ID=513, Class=3 - RPM Spoofing
344,006,028,006,063,006,042,002,041,4 => ID=344, Class=4 - SPEED Spoofing
```

**Table 3**
Two samples of raw CAN messages and their corresponding feature vectors. DLC (Data Length Code) is the 4 bits of data in the Control Field indicating the length (in these samples, 8 bytes) of each data byte. DB = Data Byte.

| CAN raw data in hexadecimal | | | Extracted feature vector in decimal | | | | | | | | |
|---|---|---|---|---|---|---|---|---|---|---|---|
| ID | DLC | Data Bytes | ID | $DB_1$ | $DB_2$ | $DB_3$ | $DB_4$ | $DB_5$ | $DB_6$ | $DB_7$ | $DB_8$ |
| 217 | 8 | 7F FF 7F FF 7F FF 7F FF | 535 | 127 | 255 | 127 | 255 | 127 | 255 | 127 | 255 |
| 123 | 8 | 09 0D 08 06 04 03 05 0E | 291 | 009 | 013 | 008 | 006 | 004 | 003 | 005 | 014 |

where $N_i$ is the number of neurons in the $i$th layer; $c$ is the total number of classes.

Here we outline the rationale behind the design and formulation of the equation presented in Eq. (1).

ANNs mimic the functioning of the human brain, where the initial layer requires a greater number of neurons or connections to effectively process and learn information. This pattern is similarly observed in ANNs, as task complexity typically diminishes from the input layer toward the output layer. The design of Eq. (1) was guided by several key considerations:

1. Based on this understanding, we aimed to allocate a higher number of neurons to the layers closer to the input.
2. To enhance the performance of multinomial classification, we sought to scale the number of neurons in each layer according to the number of classes in the dataset.
3. We also prioritized minimizing the total number of neurons due to the following factors:

    (a) The dimensionality of data and the complexity of the task, specifically, the number of input features. In the case of CAN-based attacks on IoVs, the feature count is significantly lower compared to other applications.
    (b) A larger number of neurons increases the likelihood of overfitting.
    (c) Computational constraints, including the need for real-time processing and lightweight deployment. More neurons result in greater computational demand.

Fig. 4 shows the overall architecture of the ANN described above, which forms the core of our proposed lightweight ML model. The layers are counted starting from the right, meaning from the output layer. For instance, if there are 5 classes in total, the output layer will consist of 5 neurons; the adjacent hidden layer to its left will contain 10 neurons; and this pattern continues. As a result, the input layer ends up having the highest number of neurons. The number of neurons in each layer is determined by the number of classes in the dataset and the layer's position in the count, which begins at the output layer and progresses leftward.

One of the key performance optimizations in neural networks is tuning hyperparameters such as *batch size*, number of *hidden layers*, and *epochs*. The optimal values depend on the dataset's type and complexity.

- *Batch size* refers to the number of samples processed per iteration. It can be the entire dataset or a subset, with the dataset divided into smaller batches that are propagated sequentially.
- *Hidden layers* consist of neurons that deepen the network, enabling it to learn complex data representations.

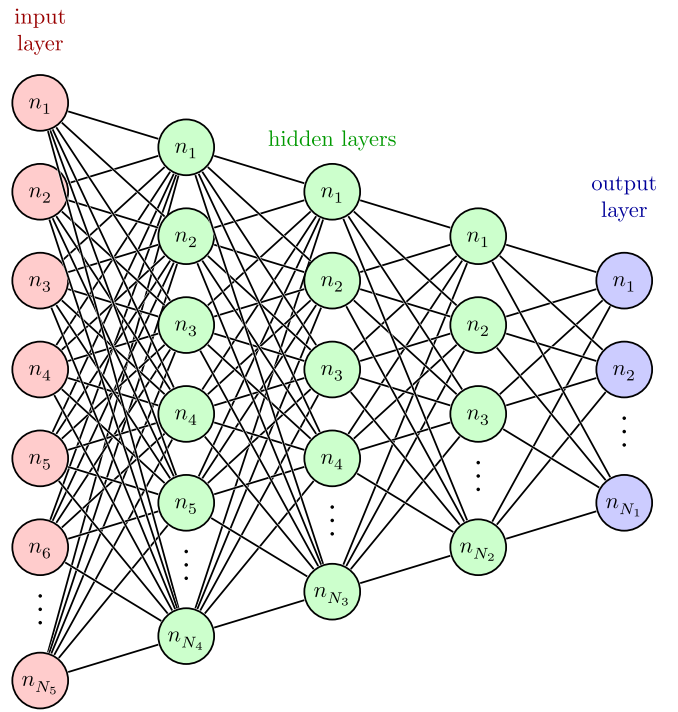

**Fig. 4.** A high-level architecture of the ANN designed and developed in the paper. $N_i = i \times c$, where $N_i =$ the number of neurons in the $i$th layer and $c =$ the total number of classes.

- *Epochs* represent the number of times the entire dataset is passed through the network during training, with values ranging from 1 – ∞.

We developed the ANN architecture by empirically determining key hyperparameters, as detailed in Section 5.3. The hyperparameters of the DAIRE model, including the number of hidden layers, batch size, and number of epochs, were determined through a structured empirical process involving random search over predefined ranges. This approach was chosen for its efficiency in exploring a broad hyperparameter space without the computational overhead of exhaustive methods like grid search. A randomly selected subset of the dataset was used for hyperparameter tuning. Various epoch values were tested, and the optimal count was selected when the model achieved an accuracy of over 99% on the subset. The following ranges were explored: number of hidden layers from 1 to 5, batch size from 100 to 500, and number of epochs from 100 to 500. A randomly selected 50% subset of the CICIoV2024 dataset was used for this tuning phase. For each combination, we performed 10-fold cross-validation to ensure robustness and avoid overfitting. Our experiments revealed that the number of hidden layers was the most sensitive hyperparameter, significantly influencing both accuracy and computational efficiency. The final selected values were then applied to train the model on the full dataset. Based on our empirical evaluations, we chose the following hyperparameter settings: $hidden\ layers = 3$; $epochs = 100$; and

$$batch\ size = \frac{training\ data\ size}{number\ of\ batches} \qquad (2)$$

where, $number\ of\ batches = 300$.

ReLU (Rectified Linear Unit) was selected as the activation function due to its computational efficiency. Other model parameters were determined accordingly based on the use of ReLU. Specifically, Kaiming initialization was adopted, as it is widely recommended for neural network layers employing ReLU activation. This initialization strategy samples initial weights from a normal distribution optimized for ReLU, helping to alleviate issues related to vanishing or exploding gradients. As a result, it enhances training stability and accelerates convergence. Additionally, L2 regularization was applied to reduce overfitting and constrain model weights, as it is a commonly used and effective regularization technique when working with ReLU-based networks.

These choices reflect a balance between model complexity and training efficiency, and they significantly contributed to the performance of our proposed approach.

### 4.3. Model training configuration

A loss function helps ML algorithms to optimize the model's predictions. They accomplish this by measuring error in the model's predictions by providing a quantitative metric for the model's accuracy. During training, the optimization algorithm uses this loss function to adjust the model's parameters to reduce the error and improve the model's predictions. For designing a lightweight ML model that classifies several different types of attacks, specifically *DoS* and various *Spoofing* attacks, we need a loss function that works well and helps in optimizing multinomial classification. For this purpose, we use the *Sparse Categorical Cross-Entropy* (SCCE) as the loss function, which is well-suited for multi-class classification with integer-encoded labels and reduces memory overhead compared to one-hot encoded alternatives (Gordon-Rodriguez et al., 2020). Unlike CCE (Categorical Cross-Entropy), which represents classes as vectors, SCCE represents classes as numbers (integers), which means it is memory and computationally efficient. This is essential when deploying models in latency-sensitive and resource-constrained environments like the IoV. Our choice to use SCCE in designing DAIRE was based on its suitability for multi-class classification, but also because of its lightweight nature. This aligns with our goal of achieving real-time detection without overloading vehicular processing units.

For optimization, we use *Root Mean Square Propagation* (RMSprop) (Hinton et al., 2012), a variant and an improvement over the standard *Gradient Descent*. RMSprop employs an adaptive learning rate method that adjusts step sizes based on recent gradient magnitudes, leading to stable and fast convergence in practice. It is computationally efficient and simpler compared to other adaptive learning methods, such as *Adam*

These choices support our objective of lightweight and real-time deployment in IoV environments, where computational efficiency and training stability are critical. The mathematical details of SCCE and RMSprop follow standard definitions and are omitted for brevity; readers are referred to the original literature for details.

## 5. Experimental evaluation

We undertook an empirical evaluation to comprehensively assess the performance of DAIRE. This section details the dataset, evaluation metrics, our empirical study, the results achieved, and the analysis. All experiments were executed on an Intel 8 Core i7-7700 equipped with 16 GB of RAM, operating on Ubuntu 24.04.

**Table 4**
Distribution of the two datasets used in this paper to evaluate our proposed model DAIRE.

| Attack type (class) | Number of CAN messages |
|---|---|
| CICIoV2024 | |
| Normal/Benign | 1,223,737 |
| DoS | 74,663 |
| Gas Spoofing | 9991 |
| RPM Spoofing | 54,900 |
| Speed Spoofing | 24,951 |
| Steering Wheel Spoofing | 19,977 |
| Total | 1,408,219 |
| Car-Hacking | |
| Normal/Benign | 928,136 |
| DoS | 587,520 |
| Fuzzy | 491,846 |
| Gear Spoofing | 597,251 |
| RPM Spoofing | 654,896 |
| Total | 3,259,649 |
| **Grand Total** | **4,667,868** |

### 5.1. Dataset

We selected two of the popular and recent IoV security datasets, CICIoV2024 (Neto et al., 2024) and Car-Hacking (Song et al., 2020), to evaluate the performance of our model. These datasets were generated on real vehicles and include *DoS, Fuzzy, RPM Spoofing, Speed Spoofing, Gas Spoofing, Gear Spoofing*, and *Steering Wheel Spoofing* attacks, and also *Attack-Free* (normal) data. The selection of these datasets was guided by their real-world fidelity, comprehensive attack coverage, and complementary characteristics that collectively enable a thorough evaluation of DAIRE. Both datasets were generated from actual vehicular systems, ensuring authentic CAN bus behavior, and include diverse attack types. While CICIoV2024 features naturally imbalanced attack frequencies that mirror realistic operational conditions, the Car-Hacking dataset provides a more balanced distribution with a larger sample size. This combination allows us to assess DAIRE's robustness across different data distributions while enabling direct comparison with state-of-the-art methods that commonly use these established benchmarks. The datasets' focus on intra-vehicle CAN attacks aligns precisely with DAIRE's threat model, making them more suitable than alternatives such as synthetically-generated datasets. Table 4 provides the distribution of these two datasets after preprocessing.

As we can see from Table 4, both datasets contain a large number of samples in each class, which is suitable for training our proposed lightweight ML model DAIRE. The Car-Hacking dataset is more balanced than CICIoV2024. Using the Pareto principle (Pareto, 1919), for evaluating DAIRE, we partitioned the dataset into 80% training and 20% testing. For tuning DAIRE's hyperparameters, we used the newer CICIoV2024 dataset and randomly chose 50% of the samples from each class for tuning these parameters.

### 5.2. Evaluation metrics

The lightweight ML model DAIRE proposed in this paper learns from the previous attacks and then detects and predicts the class of the new attack against IoV. DAIRE performs multinomial classification to detect and classify such attacks. One of the most commonly used methods for evaluating and visualizing the performance of such a classifier is the confusion matrix (Fawcett, 2006). A confusion matrix serves as a powerful tool for evaluating and visualizing the performance of a classifier. By leveraging this matrix, we gain crucial insights into the classifier's strengths and weaknesses, paving the way for targeted improvements. Moreover, the confusion matrix enables us to calculate essential evaluation metrics, ensuring our classifier operates at its highest potential. An example of a confusion matrix of three classes A, B, and C is shown in Table 5.

**Table 5**
An example of a simple confusion matrix of three classes.

|   | A | B | C |
|---|---|---|---|
| A | $TP_{AA}$ | $F_{AB}$ | $F_{AC}$ |
| B | $F_{BA}$ | $TP_{BB}$ | $F_{BC}$ |
| C | $F_{CA}$ | $F_{CB}$ | $TP_{CC}$ |

The rows of a confusion matrix represent the actual class of samples, while the columns represent the predicted class. All the correct predictions are on the diagonal of the confusion matrix (highlighted in green).

For class *A* we compute:

- *True positives* (TP) ⇒ $TP_{AA}$, i.e., samples labeled as class *A* correctly classified as *A*.
- *True Negatives* (TN) ⇒ $TN_A = TP_{BB} + TP_{CC} + F_{BC} + F_{CB}$, i.e., samples labeled as other classes, in this case classes *B* and *C* are predicted as either *B* or *C*.
- *False Positives* (FP) ⇒ $FP_A = F_{BA} + F_{CA}$, i.e., samples labeled as other classes, in this case classes *B* and *C*, are predicted as class *A*.
- *False Negatives* (FN) ⇒ $FN_A = F_{AB} + F_{AC}$, i.e., samples labeled as class *A* predicted as other classes, in this case classes *B* and *C*.

Similarly, we can compute TP, TN, FP, and FN for other classes.

Using these notations from the confusion matrix, we compute the following metrics to evaluate the performance of DAIRE.

The *Detection Rate (DR)*, also known as *Recall* or *True Positive Rate*, measures the number of true samples that have been accurately predicted as true. It is defined as follows:

$$DR = \frac{TP}{TP + FN}$$

*False Positive Rate* (FPR) refers to the number of negative samples that have been incorrectly predicted as true. It is defined as:

$$FPR = \frac{FP}{TN + FP}$$

*Accuracy* measures the total number of samples that have been correctly predicted. It is defined as:

$$Accuracy = \frac{TP + TN}{TP + TN + FP + FN}$$

*Precision* indicates the proportion of samples that have been predicted as true, regardless of whether these samples are actually true or false. It is defined as:

$$Precision = \frac{TP}{TP + FP}$$

$F_1$-score is the weighted average of precision and recall. It is defined as:

$$F_1 - score = \frac{2TP}{2TP + FP + FN}$$

### 5.3. Experiments and results

We conducted a series of systematic experiments to evaluate and validate the effectiveness and efficiency of the proposed DAIRE model. The experiments were performed using the two recent and widely used IoV datasets: CICIoV2024 and Car-Hacking. Our experiments were designed to (i) tune the hyperparameters, (ii) evaluate model performance under real-world conditions, and (iii) measure classification accuracy, speed, and resource usage.

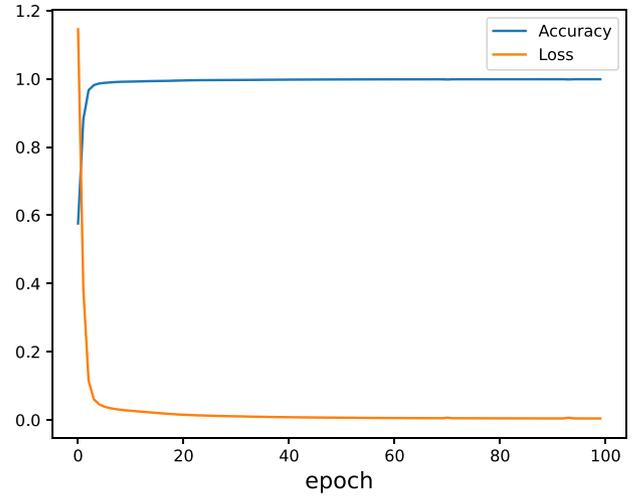

**Fig. 5.** Accuracy and loss plot during one of the training runs of DAIRE for tuning of the hyperparameters. DAIRE achieved an accuracy of 99.9% and a loss of 0.0045 at 100 count of epochs.

#### 5.3.1. Hyperparameter tuning on CICIoV2024 subset

In the first experiment, we focused on determining the optimal hyperparameters for the DAIRE model. For this experiment, we randomly chose 50% of the samples of each class from the CICIoV2024 dataset. We then partitioned the dataset into 80% training and 20% testing sets. We ran the experiment and completed 10 iterations each time using a different set of 20% for testing. The results obtained from these experiments are shown in Table 6 and visualized using a confusion matrix in Fig. 7 – 7(e). The *number of batches* of 300 in Eq. (2) and *epochs* of 100 were computed empirically, conducting similar experiments each time with different values of the number of batches and epochs. Based on our previous experience (Alam & Demir, 2024) of designing and developing lightweight ML models, we tried values in the range from 100–500 for selecting the best of these two parameters. We employed a random search methodology for hyperparameter optimization, exploring the following ranges: number of hidden layers from 1 to 5, batch size from 100 to 500, and number of epochs from 100 to 500. A randomly selected 50% subset of the CICIoV2024 dataset was used for this tuning phase. For each combination, we performed 10-fold cross-validation to ensure robustness and avoid overfitting. The optimal configuration was found to be: hidden layers = 3, batch size = training data size/300, and epochs = 100. This configuration achieved the best balance between high detection accuracy and low computational latency. The experiments with different hidden layers in the range from 1 – 5 are shown in Table 6. DAIRE achieved the best results when 3 hidden layers were used. Fig. 5 shows the accuracy and loss plot during training and tuning of DAIRE when run with these values. Fig. 6 visually summarizes the average classification performance across different hidden layer configurations, demonstrating that DAIRE achieves optimal results with 3 hidden layers. The consistent improvement in DT and reduction in false positive rates as the number of hidden layers increases from 1 to 3 validates our architectural design choices. Based on these experiments, we designed DAIRE with *hidden layers* = 3, *number of batches* = 300, and *epochs* = 100, and used the same number of parameters for the rest of the experiments in this paper.

#### 5.3.2. Evaluation on CICIoV2024 dataset

Using the optimal hyperparameters, we evaluated DAIRE on the complete CICIoV2024 dataset as listed in Table 4 with a total number of 1,408,219 CAN messages across six classes. It was partitioned into 80% training and 20% testing. We ran the experiment and completed 10 iterations each time using a different set of 20% for testing. The results obtained are shown in Table 7 and visualized using a confusion matrix in Fig. 7(c).

**Table 6**
Classification results of different classes by DAIRE on the 50% randomly chosen samples from the CICIoV2024 dataset using different *hidden layers* with *batch size* = $\frac{training\ data\ size}{300}$, and *epochs* = 100. An ↑ signifies that a higher value is preferable, while a ↓ indicates that a lower value is preferable.

| Attack Type | DR ↑ | FPR ↓ | Accuracy ↑ | Precision ↑ | $F_1$-score ↑ |
|---|---|---|---|---|---|
| *hidden layer = 1* | | | | | |
| Normal/Benign | 99.05% | 4.2% | 98.62% | 99.21% | 99.36% |
| DoS | 100% | 0.2% | 99.77% | 97.87% | 95.86% |
| Gas Spoofing | 100% | 0% | 100% | 100% | 100% |
| RPM Spoofing | 99.99% | 0.5% | 99.47% | 93.65% | 88.06% |
| Speed Spoofing | 69.45% | 0% | 99.45% | 81.75% | 99.35% |
| Steering Wheel Spoofing | 99.95% | 0.05% | 99.94% | 97.93% | 95.99% |
| Average | 94.74% | 0.8% | 99.54% | 95.07% | 96.45% |
| *hidden layers = 2* | | | | | |
| Normal/Benign | 99.85% | 0% | 99.87% | 99.92% | 99.99% |
| DoS | 100% | 0% | 99.99% | 99.97% | 99.94% |
| Gas Spoofing | 100% | 0% | 100% | 100% | 100% |
| RPM Spoofing | 99.99% | 0.5% | 99.99% | 99.93% | 99.87% |
| Speed Spoofing | 100% | 0.1% | 99.88% | 96.69% | 93.59% |
| Steering Wheel Spoofing | 99.95% | 0% | 99.99% | 99.77% | 100% |
| Average | 99.96% | 0.02% | 99.96% | 99.41% | 98.90% |
| *hidden layers = 3* | | | | | |
| Normal/Benign | 99.99% | 0% | 99.99% | 99.99% | 99.99% |
| DoS | 100% | 0% | 99.99% | 99.97% | 99.94% |
| Gas Spoofing | 100% | 0% | 100% | 100% | 100% |
| RPM Spoofing | 99.99% | 0% | 99.99% | 99.99% | 100% |
| Speed Spoofing | 100% | 0% | 100% | 100% | 100% |
| Steering Wheel Spoofing | 99.95% | 0% | 99.94% | 99.97% | 100% |
| **Average** | **99.99%** | **0%** | **99.99%** | **99.99%** | **99.99%** |
| *hidden layers = 4* | | | | | |
| Normal/Benign | 99.99% | 0% | 99.99% | 99.99% | 99.99% |
| DoS | 100% | 0% | 99.99% | 99.99% | 99.99% |
| Gas Spoofing | 100% | 0% | 100% | 100% | 100% |
| RPM Spoofing | 99.99% | 0% | 99.99% | 99.94% | 99.89% |
| Speed Spoofing | 100% | 0% | 99.99% | 99.96% | 99.92% |
| Steering Wheel Spoofing | 99.95% | 0% | 99.99% | 99.97% | 100% |
| Average | 99.99% | 0% | 99.99% | 99.98% | 99.97% |
| *hidden layers = 5* | | | | | |
| Normal/Benign | 99.99% | 0% | 99.99% | 99.99% | 99.9% |
| DoS | 100% | 0% | 99.99% | 99.96% | 99.93% |
| Gas Spoofing | 100% | 0% | 99.99% | 99.89% | 99.79% |
| RPM Spoofing | 99.99% | 0% | 99.99% | 99.97% | 99.95% |
| Speed Spoofing | 100% | 0% | 99.99% | 99.97% | 99.94% |
| Steering Wheel Spoofing | 99.95% | 0% | 99.99% | 99.97% | 100% |
| Average | 99.99% | 0% | 99.99% | 99.96% | 99.94% |

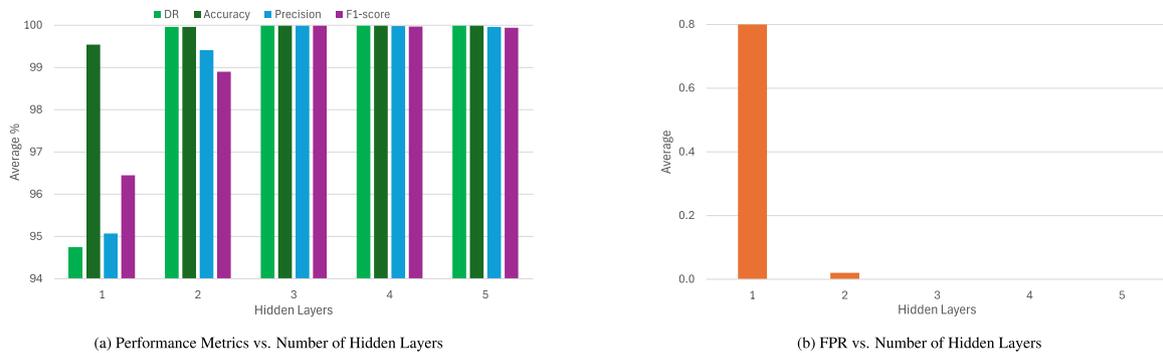

(a) Performance Metrics vs. Number of Hidden Layers

(b) FPR vs. Number of Hidden Layers

**Fig. 6.** Average classification results by DAIRE on the 50% randomly chosen samples from the CICIoV2024 dataset using different *hidden layers* with *batch size* = $\frac{training\ data\ size}{300}$, and *epochs* = 100.

*5.3.3. Evaluation on the car-hacking dataset*

To evaluate generalizability, we tested DAIRE on a larger dataset, i.e., the Car-Hacking dataset, which contains over 3.2 million CAN messages. For this experiment, we used the complete Car-Hacking dataset as listed in Table 4. It was partitioned into 80% training and 20% testing. We ran the experiment and completed 10 iterations each time using a different set of 20% for testing, also known as 10-fold cross-validation. The results obtained are shown in Table 7 and visualized using a confusion matrix in Fig. 7(f).

*5.3.4. Results analysis*

Fig. 8 presents the Coefficient of Variation (CV) (Brown, 1998) values obtained from 10-fold cross-validation for each attack type across both the datasets (CICIoV2024 and Car-Hacking). The CV reflects the degree of variability or instability in a model's performance over different random data partitions and is computed as (Standard Deviation of Scores/Mean Score) × 100%. The observed CV values (Fig. 8) range between 0% and 3.78%, indicating minimal variance and demonstrating that DAIRE maintains highly consistent performance across varying data subsets.

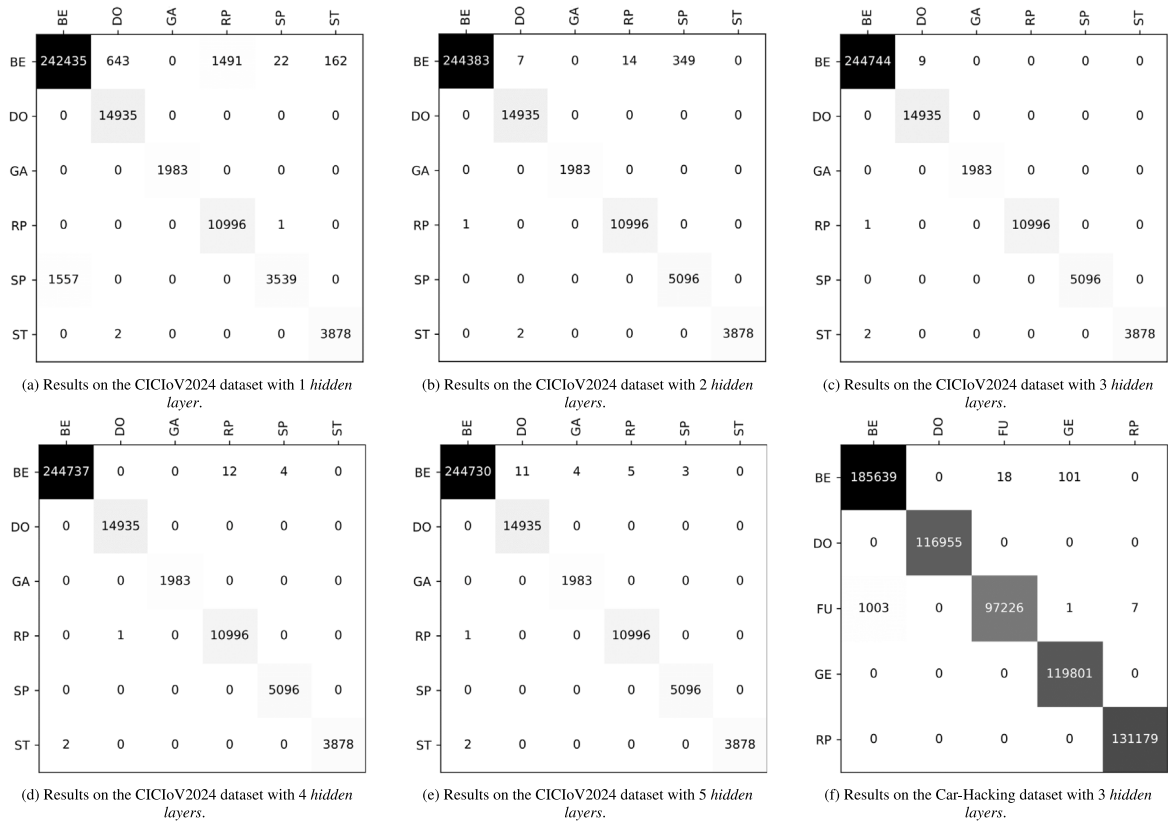

Fig. 7. Confusion matrix visualization of the results provided in Tables 6 and 7 on the two datasets using different *hidden layers* with *batch size* = $\frac{training\ data\ size}{300}$ and *epochs* = 100. BE = Benign, DO = DoS, GA = Gas, RP = RPM, SP = Speed, ST = Steering Wheel, GE = Gear, and FU = Fuzzy.

**Table 7**
Classification results of different classes by DAIRE on the two datasets with *hidden layers* = 3, *batch size* = $\frac{training\ data\ size}{300}$, and *epochs* = 100. An ↑ signifies that a higher value is preferable, while a ↓ indicates that a lower value is preferable.

| Attack Type | DR ↑ | FPR ↓ | Accuracy ↑ | Precision ↑ | $F_1$-score ↑ |
|---|---|---|---|---|---|
| CICIoV2024 | | | | | |
| Normal/Benign | 99.99% | 0% | 99.99% | 99.99% | 99.99% |
| DoS | 100% | 0% | 99.99% | 99.97% | 99.94% |
| Gas Spoofing | 100% | 0% | 100% | 100% | 100% |
| RPM Spoofing | 99.99% | 0% | 99.99% | 99.99% | 100% |
| Speed Spoofing | 100% | 0% | 100% | 100% | 100% |
| Steering Wheel Spoofing | 99.95% | 0% | 99.94% | 99.97% | 100% |
| Average | 99.99% | 0% | 99.99% | 99.99% | 99.99% |
| Car-Hacking | | | | | |
| Normal/Benign | 99.94% | 0.2% | 99.83% | 99.70% | 99.46% |
| DoS | 100% | 0% | 100% | 100% | 100% |
| Fuzzy | 98.97% | 0% | 99.84% | 99.47% | 99.98% |
| Gear Spoofing | 100% | 0% | 99.98% | 99.96% | 99.91% |
| RPM Spoofing | 100% | 0% | 99.99% | 99.99% | 99.99% |
| Average | 99.78% | 0.04% | 99.93% | 99.83% | 99.87% |

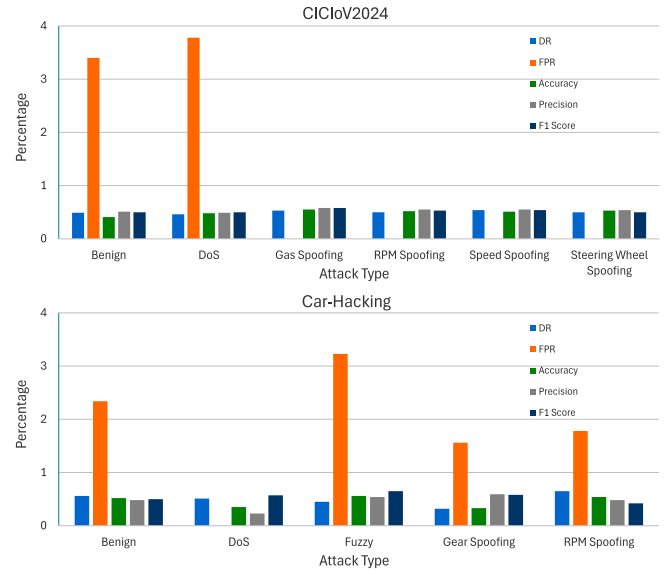

Fig. 8. Coefficient of Variation values (in percentage) obtained from 10-fold cross-validation for each attack type across both the datasets, CICIoV2024 and Car-Hacking.

The results presented in Table 7 show that DAIRE performs slightly better on the CICIoV2024 dataset compared to the Car-Hacking dataset. This difference in performance can be attributed to the hyperparameters of DAIRE being specifically tuned for the CICIoV2024 dataset.

When tested on the larger (with a total number of 3,259,649 CAN messages) dataset Car-Hacking, DAIRE achieved an average DR of 99.78% and a FPR of 0.04% with an accuracy of 99.93%. On average for the two datasets, DAIRE achieved a DR of 99.88% and a FPR of 0.02% with an accuracy of 99.96%. These metrics demonstrate DAIRE's exceptional capability in detecting and classifying various IoV attacks.

**Table 8**
Results of the experiments on CICIoV2024 dataset to evaluate the performance of our allocation scheme given in Eq. (1), comparing it with varying fixed neuron counts. An ↑ signifies that a higher value is preferable, while a ↓ indicates that a lower value is preferable.

| Number of neurons in each layer | Training time ↓ | Testing time ↓ | TPR ↑ | FPR ↓ |
| --- | --- | --- | --- | --- |
| $i \times c$ (Eq. (1)) | **80.69 s** | **10.1 s** | **99.998%** | **0.00000452** |
| 10 | 98.9 s | 11.1 s | 99.989% | 0.00001360 |
| 25 | 150.1 s | 11.6 s | 99.989% | 0.00001360 |
| 50 | 165.86 s | 12.2 s | 99.998% | 0.00000452 |
| 100 | 236.26 s | 12.7 s | 99.998% | 0.00000452 |
| 200 | 452.83 s | 13.1 s | 99.998% | 0.00000452 |

By carefully selecting and optimizing various components of our model, DAIRE achieved a classification time of just 0.03 ms per sample and consumed 0.35 KB of memory when evaluated on the Car-Hacking dataset. This demonstrates DAIRE's exceptional capability to classify IoV attacks in real-time.

A critical observation from our results is DAIRE's consistently high performance across both majority and minority classes, despite a significant imbalance in the CICIoV2024 dataset. This robustness can be attributed to several architectural and methodological factors. First, spoofing attacks such as Gas Spoofing and Steering Wheel Spoofing inject messages with distinctive features that create strong signals detectable even with limited samples. The Sparse Categorical Cross-Entropy loss function ensures that misclassifications on minority classes receive an appropriate penalty during training, preventing the model from being biased toward the predominant *Normal* class. Finally, the empirical design of our ANN architecture, with its controlled capacity via the neuron allocation formula, provides sufficient representational power to capture minority class patterns without overfitting. These design choices collectively enable DAIRE to achieve balanced performance across all attack categories, demonstrating its suitability for real-world IoV environments where attack frequencies are inherently imbalanced.

Although DAIRE's lightweight design and CAN-specific constraints, including message formats and permissible value ranges, may limit the feasibility of adversarial attacks, attackers familiar with CAN could still craft convincing injections or alter message sequences. Thus, DAIRE's robustness against adversarial scenarios cannot be taken for granted. In future work, we plan to experimentally assess DAIRE's resilience under diverse attacker models and present both success and failure rates across different perturbation strategies.

*5.3.5. Evaluating the allocation scheme for neurons (equ (1))*

We conducted a performance evaluation of our allocation scheme for neurons given in Eq. (1). For this, we ran a series of similar experiments as described in Section 5.3.2, comparing the allocation scheme against standard configurations with varying fixed neuron counts (each layer with a fixed number of neurons). The results are shown in Table 8.

The experiments were conducted using the CICIoV2024 dataset, with the reported time representing the total training and testing duration. The first row in the table presents the performance of our allocation scheme. The FPR values are nearly zero but are included for comparison since they differ slightly. Although increasing the number of neurons enhances TPR and FPR, it also results in longer training and testing times. In contrast, our allocation scheme achieves the same TPR and FPR as configurations with 50, 100, and 200 neurons, while significantly reducing both training and testing time. These findings further reinforce and validate the formulation of Eq. (1) for selecting the optimal number of neurons in each layer, enabling real-time processing and lightweight deployment.

*5.3.6. Interpretability analysis*

Although DAIRE is implemented as a lightweight feed-forward ANN, its ability to distinguish between different spoofing attacks is explained by the structured nature of CAN message encoding. Each vehicle function (e.g., speed, engine RPM, throttle position, gear state) is represented by fixed Arbitration IDs and specific payload byte positions. Spoofing attacks typically manipulate only those CAN fields related to the targeted function, leading to consistent, class-specific statistical deviations in a limited subset of input features. For example, speed spoofing primarily alters payload bytes corresponding to vehicle velocity signals, whereas RPM spoofing affects different byte fields linked to engine rotation. Fuzzy attacks, in contrast, inject random or malformed values across multiple identifiers and payload bytes, producing high-variance distributions that are distinct from structured spoofing patterns. Because these deviations are systematic and repeatable, the ANN learns stable decision boundaries based on correlations among specific identifiers and payload bytes. This is the reason why the proposed architecture is sufficient to achieve strong multi-class discrimination across spoofing categories.

## 6. Discussion

This section presents a comparison between DAIRE and six closely related state-of-the-art approaches, while also outlining potential future directions to enhance attack classification in IoV.

*6.1. Comparison with state-of-the-art works*

For a general comparison of DAIRE with other works, a literature review summary of recent works on IoV attack detection is presented in Table 1. In this section, we specifically compare DAIRE with six state-of-the-art works. All these six works report the classification (testing) time; two of them use the same dataset, and all of them employ an ML model to detect/classify IoV attacks. Table 9 presents a comparison of DAIRE with these six works. Fig. 9 visually summarizes the comparison. DAIRE achieves competitive DR while maintaining significantly faster classification times than all compared state-of-the-art approaches, validating its suitability for real-time IoV deployment.

In IoV intrusion detection systems, the need for real-time responsiveness usually necessitates that each data packet be analyzed in under 10 ms (ms) (Moubayed et al., 2020). All the works compared satisfy this requirement. Here we would like to highlight DAIRE's superior ability to classify IoV attacks in real-time due to its better latency and smaller memory footprint than all the other six works.

By specifically choosing and optimizing the different components of our model, DAIRE achieved a classification time of 0.03 ms per sample (input). Although Song et al. (Song et al., 2020) are using a deep CNN model but they still achieve a much better time of 0.15 ms per sample than others. They are using one of the CNN models, Reduced Inception-ResNet, which consists of multiple sets of convolutional layers, and that may be one of the reasons their classification time is larger than DAIRE. Also, their memory consumption is larger because they convert CAN data into images. Yang et al. (2021) and Qin et al. (2024) also achieve a classification time < 1 ms, but the memory used by Yang et al. (2021) is much larger than DAIRE (because of stacking and clustering), and the use of a proprietary dataset in Qin et al. (2024) makes it difficult to compare with DAIRE.

Although Huang et al. (2024) employs a lightweight CNN and has eliminated unnecessary layers, CNNs inherently include multiple convolutional layers and are designed to process images. As a result, their classification time remains significantly higher than that of DAIRE. For the same reasons, even employing a lightweight CNN, the classification time of Alladi et al. (2021) is > 1 ms.

While DAIRE was evaluated on a more powerful machine compared to some of the other works (using ARM Cortex), the disparity in

**Table 9**
Comparison of DAIRE with six state-of-the-art works. CTS = Classification time per sample. MUS = Memory used (training + classification) per sample. NR = Not reported. For a fair comparison, the results reported here for DAIRE are only for the Car-Hacking dataset. An ↑ signifies that a higher value is preferable, while a ↓ indicates that a lower value is preferable.

| Model | DR ↑ | FPR ↓ | CTS ↓ | MUS ↓ | Dataset | Testing device | Methodology |
|---|---|---|---|---|---|---|---|
| **DAIRE** | 99.78% | 0.04% | 0.03 ms | 0.35 KB | Car-Hacking | Intel i7-7700; 3.6 GHz; single core. | CAN data to decimal; Lightweight ANN. |
| Song et al. (2020) | 99.84% | NR | 0.15 ms | 2.2 MB | Car-Hacking | Intel Xeon; 2.3 GHz; single core. | CAN data to image; Deep CNN. |
| Huang et al. (2024) | 99.21% | NR | 6 ms | NR | Car-Hacking | ARM Cortex-A72; 1.8 GHz; single core. | CAN data to image; Lightweight CNN. |
| Yang et al. (2021) | 99.88% | 0.05% | 0.54 ms | 18.8 MB | CAN-Intrusion; CICIDS-2017 | ARM Cortex-A53; upto 1.4 GHz; single core. | Multitiered; Stacking; k-means clustering. |
| Qin et al. (2024) | 99.61% | 5.95% | 0.1 ms | NR | Proprietary | Intel Core i5; upto 4.5 GHz; single core. | Cloud-based training; XGBoost. |
| Chen et al. (2024) | 98.37% | NR | 5-9 ms | NR | CAN-Intrusion; UNSWNB-15; CICIDS-2017 | ARM Cortex-A72; 1.5 GHz; single core. | Federated learning; Differential privacy. |
| Alladi et al. (2021) | 98.26% | NR | 1.55 ms | NR | VeReMi Extension | ARM Cortex-A53; 1.2 GHz; single core. | CAN data to image; Lightweight CNN. |

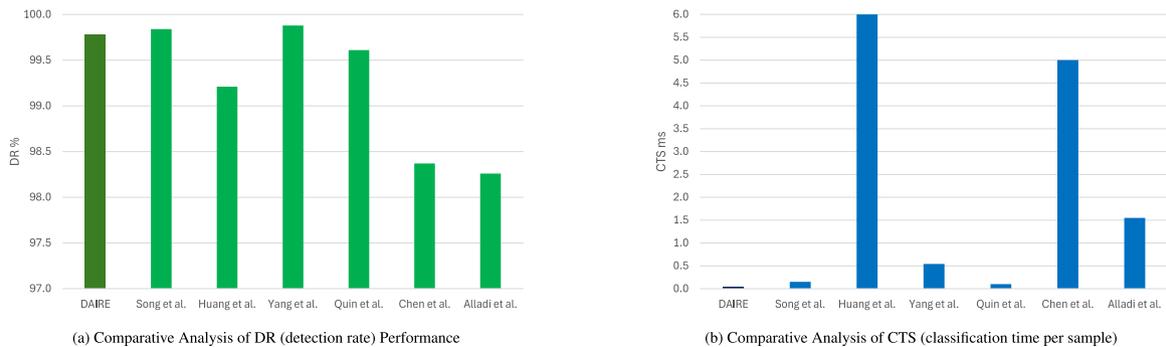

(a) Comparative Analysis of DR (detection rate) Performance

(b) Comparative Analysis of CTS (classification time per sample)

**Fig. 9.** Comparison of DAIRE with six state-of-the-art works.

computational power alone does not account for the considerable time difference observed.

DAIRE achieves an average DR of 99.78% on the Car-Hacking dataset, performing on par with or better than the six other studies. Its FPR of 0.04% surpasses the two other works that reported their FPRs. Our lightweight ANN model is developed through an empirical design approach, with a detailed explanation of the optimization process and how it was tailored to the specific task using experiments conducted on a subset of the dataset.

*6.2. Future works*

There are two deployment approaches for an IDS for a network of IoVs. Centralized and distributed (Taslimasa et al., 2023). In a centralized approach, the IDS is deployed on a remote cloud or an edge server. In a distributed approach, the IDS is deployed on each network node, i.e., an IoV, and the attack data is shared collaboratively. There are pros and cons in both approaches. A distributed approach is more secure and private and does not require high bandwidth. Whereas, centralized approach is more suitable for resource-constrained IoVs. Some disadvantages of a distributed approach are: lack of accuracy; heterogeneous network with different normal and abnormal behaviors; and the collaboration results may diverge from the global optimized point. Some disadvantages of a centralized approach are: the single point of failure, scalability, and the requirement of a large bandwidth. *IDS deployment* is an open research problem (Taslimasa et al., 2023). In the future, we would like to explore the installation and deployment of DAIRE as part of an IDS using a distributed approach on a network of IoVs using TinyML (Ray, 2022). In our opinion distributed approach is more scalable and secure.

DAIRE is evaluated using datasets for wired networks that contain specific attacks, such as DoS, fuzzy, and spoofing. In the future, we would like to evaluate DAIRE against other types of *generalized datasets* that also contain malicious activities related to mobility features and data with varying traffic conditions, such as routing attacks, speed of vehicle, and distance between vehicles, etc.

In the future, we aim to evaluate DAIRE against adversarial ML attacks (Malik et al., 2024). These attacks are designed to disrupt the prediction accuracy of an ML model by altering inputs, leading the model to misclassify. If successfully executed, they can greatly diminish a model's precision and in the case of IoVs, can lead to severe damage. The Adversarial Robustness Toolbox (ART) (Nicolae et al., 2018) is an open-source Python library developed by IBM Research. It is designed to help researchers and practitioners assess and enhance the resilience of ML models against adversarial attacks. We will use ART to create adversarial examples of IoV attacks and evaluate DAIRE's performance in response to these attacks.

*6.3. Real-world deployment considerations*

DAIRE is capable of detecting not only major types of CAN attacks but also less common attacks, such as Gas Spoofing and RPM Spoofing, which are crucial for real-world applications. With an inference time of just 0.03 ms, DAIRE offers deterministic latency, making it suitable for real-time operation on modern embedded systems. Two well-known vendors of AI for embedded systems, Raspberry Pi[1] and Nvidia Jetson,[2] utilize ARM CPUs, specifically the Armv8 (e.g., Cortex-A78) and Armv9 (e.g., Cortex-X925). These CPUs can achieve clock speeds of up to 3.0 GHz to 3.8 GHz and can be integrated with sufficient RAM to support lightweight AI models like DAIRE. With moderate updates, DAIRE can be effectively deployed on modern embedded systems designed for lightweight AI applications. To validate this belief, we plan to conduct tests for deploying DAIRE on platforms such as Raspberry Pi and Nvidia Jetson in the future.

## 7. Conclusion

The rapid advancement of the Internet of Vehicles (IoV) opens up significant opportunities while simultaneously introducing critical security challenges. A primary concern lies in the dependence on the CAN protocol, which renders vehicular systems susceptible to serious cyber threats due to its lack of built-in security mechanisms. In response to this pressing issue, we proposed DAIRE, a lightweight machine learning-based IDS specifically designed for real-time attack detection and classification within CAN-enabled IoV environments. DAIRE employs a lightweight ANN architecture, thereby making it well-suited for deployment in resource-constrained vehicular systems. By integrating SCCE and RMSprop optimization techniques, the model not only achieves high detection performance but also maintains low computational overhead. Comprehensive evaluations conducted on benchmark datasets, including CICIoV2024 and Car-Hacking, substantiate DAIRE's robust performance and generalizability across a diverse array of attack types. The findings further demonstrate its superior real-time responsiveness compared to existing state-of-the-art methods, underscoring its practicality for enhancing the security of modern IoV systems.

**Funding**

This research is funded by the Scientific Research Deanship at the University of Ha'il, Saudi Arabia, under Grant RG-25 030.

**Declaration of competing interest**

The authors declare that they have no known competing financial interests or personal relationships that could have appeared to influence the work reported in this paper.

**Data availability**

The data utilized in this paper is publicly available from Neto et al. (2024), Song et al. (2020). The code developed in this paper is available on request.

---

[1] https://www.raspberrypi.com/
[2] https://www.nvidia.com/en-us/autonomous-machines/embedded-systems/